\documentclass[twocolumn,amsfonts,showpacs,superscriptaddress,nofootinbib]{revtex4-1}
\usepackage[dvipdfmx]{graphicx}
\usepackage{amssymb,amsmath,amsthm,booktabs,mathtools}
\usepackage{bm}
\usepackage{color}

\newcommand{\bc}{\begin{center}}
\newcommand{\ec}{\end{center}}
\def\ba#1{\begin{array}{#1}\displaystyle}
\newcommand{\ea}{\end{array}}

\newcommand{\beq}{\begin{equation}}
\newcommand{\eeq}{\end{equation}}
\newcommand{\beqa}{\begin{eqnarray}}
\newcommand{\eeqa}{\end{eqnarray}}

\newcommand{\bi}{\begin{itemize}}
\newcommand{\ei}{\end{itemize}}

\def\lt#1{\left#1}
\def\rt#1{\right#1}

\def\frc#1#2{\frac{#1}{#2}}

\newcommand{\p}{\partial}

\newcommand{\bra}{\langle}
\newcommand{\ket}{\rangle}

\newcommand{\R}{{\mathbb{R}}}

\newcommand{\Or}{{\cal O}}

\newcommand{\ep}{\epsilon}

\newcommand{\ri}{{\rm i}}
\newcommand{\dd}{{\rm d}}

\newcommand{\dr}{\mathrm{dr}}

\newcommand{\F}{{\rm F}}

\def\eqref#1{(\ref{#1})}

\usepackage{amsmath}	
\begin{document}


\title{Large-scale description of interacting one-dimensional Bose gases:\\ generalized hydrodynamics supersedes conventional hydrodynamics}

\author{Benjamin Doyon}
\affiliation
{Department of Mathematics, King's College London, Strand, London WC2R 2LS, UK
}
\author{J\'er\^ome Dubail}
\affiliation
{CNRS \& IJL-UMR 7198, Universit\'e de Lorraine, F-54506 Vandoeuvre-l\`es-Nancy, France
}
\author{Robert Konik}
\affiliation
{Condensed Matter and Materials Science Division, Brookhaven National Laboratory, Upton, NY 11973 USA}

\author{Takato Yoshimura}
\affiliation
{Department of Mathematics, King's College London, Strand, London WC2R 2LS, UK
}
\begin{abstract}
The theory of generalized hydrodynamics (GHD) was recently developed as a new tool for the study of inhomogeneous time evolution in many-body interacting systems with infinitely many conserved charges. 
In this letter, we show that it supersedes the widely used conventional hydrodynamics (CHD) of one-dimensional Bose gases. We illustrate this by studying ``nonlinear sound waves" emanating from initial density accumulations in the Lieb-Liniger model. We show that, at zero temperature and in the absence of shocks, GHD reduces to CHD, thus for the first time justifying its use from purely hydrodynamic principles. We show that sharp profiles, which appear in finite times in CHD, immediately dissolve into a higher hierarchy of reductions of GHD, with no sustained shock. CHD thereon fails to capture the correct hydrodynamics. We establish the correct hydrodynamic equations, which are finite-dimensional reductions of GHD characterized by multiple, disjoint Fermi seas. We further verify that at nonzero temperature, CHD fails at all nonzero times. Finally, we numerically confirm the emergence of hydrodynamics at zero temperature by comparing its predictions with a full quantum simulation performed using the NRG-TSA-ABACUS algorithm. The analysis is performed in the full interaction range, and is not restricted to either weak- or strong-repulsion regimes.
\end{abstract}

\maketitle
\noindent {\bf\em Introduction.}\quad 
Modern experiments with ultracold atoms confined in ``cigar-shaped'' traps \cite{coldatoms,qnc} or in atom chips \cite{chips} provide real-world implementations of one-dimensional (1d) many-body systems \cite{giamarchi}, and represent an important challenge for theoretical physics. Even though it is widely accepted that 1d clouds of bosonic atoms are described, at the microscopic scale, by the paradigmatic Lieb-Liniger (LL) model \cite{olshanii,review_giamarchi}, solving this model in experimentally relevant out-of-equilibrium inhomogeneous situations for more than a few dozens of atoms is a task that is out of the reach of modern theoretical methods, including state-of-the-art numerical ones.

Yet, it is a classic result of XX$^{\rm th}$ century mathematical physics that, in its homogeneous, translation-invariant version at equilibrium, the LL model is exactly solvable by means of the Bethe ansatz \cite{LiebLiniger}, and its equation of state can be calculated exactly \cite{yangyang}. It then seems reasonable to use this equation of state as the basic input into a coarse-grained, hydrodynamic, approach, that is expected to be applicable as soon as typical lengths of variations of the local density are large enough as compared to inter-particle and scattering distances (the Euler scale) -- in much the same way that classical hydrodynamics describes water waves. Such a "conventional hydrodynamic" (CHD) approach -- defined in Eqs. (\ref{chd}) below --, has been used extensively in the cold atoms literature over the past decade \cite{chdpapers,damski,pv14}, and has sometimes been viewed as a consequence of the Gross-Pitaevskii equation \cite{review_giamarchi,pv14} in the regime of small interaction strength.

However, a key physical feature of the LL model is overlooked in CHD: the fact that it admits infinitely many conservation laws. Indeed, CHD focuses only on a few quantities, like the particle density, the momentum density or the energy density. But the LL model possesses infinitely more conserved quantities. Those are not just a mathematical curiosity:  they can have dramatic physical consequences, as illustrated by the quantum Newton cradle experiment \cite{qnc}: the crucial observation of undamped oscillations in this experiment is connected with the lack of conventional thermalization \cite{thermalization}.


The full connection between generalized thermalization and many-body dynamics was only recently uncovered \cite{ghd,bertini1}.
The fundamental precepts of hydrodynamics -- the emergence of local entropy maximization -- were used in systems with an infinite number of conservation laws in order to form the theory of generalized hydrodynamics (GHD). 
It is a type of Euler-scale hydrodynamics, but with an infinite-dimensional space of fluid states accounting for the large state manifold accessible by generalized thermalization. 
In practice, GHD consists in an infinite set of coupled continuity equations (one for each conserved charge), that can, at least in principle, be worked out with numerical solvers for non-linear partial differential equations.

In this letter, we show that GHD supersedes CHD. 
For this purpose, we focus on far-from-equilibrium waves emanating from a density accumulation in the LL model. The density waves are a good illustration for our purposes, but the main results and mechanisms are general. They have been studied in several ways in the past decade in the free Fermi gas \cite{bumps_free_fermions}, in the effective theory of the non-linear Luttinger liquid \cite{bumps_NLLL}, and in the Calogero-Sutherland model \cite{bumps_Calogero}, quantum Hall edges \cite{bumps_FQH}, and the LL model \cite{damski} using CHD. In particular, all these references -- see also \cite{pv14} -- pointed out that the applicability of CHD was limited by the appearance of shocks. In this Letter, we show that GHD is the proper hydrodynamic framework to go beyond the latter.

We demonstrate that, only at zero temperature and for finite evolution times does CHD coincide with GHD. 
CHD being a finite-component  hydrodynamics, it inevitably leads to ``gradient catastrophes'' and shock propagations thereon. In contrast, we show that at zero entropy, GHD decomposes into a hierarchy of instantaneously invariant finite-dimensional subspaces, whose exact hydrodynamic equations we establish. These are described by a multitude of Fermi seas,
the stability of which is a consequence of integrability. We show that shocks dissolve as the system leaves the CHD subspace into a higher-dimensional reduction of GHD.
No shock propagates in this process, as instead smoothness is re-established. We note that an important practical consequence of the zero-entropy reduction is that the infinite system of coupled non-linear equations of GHD collapses to a finite number of equations that are computationally easy to solve, taking typically a few minutes on a laptop. We also numerically verify that at nonzero temperature, the GHD evolution, which necessitates the full infinite-dimensional space, is different from CHD at all times. In the density wave problem, a stark difference is that no sharp profile develops in GHD, while CHD based on the finite-temperature LL equations of state has gradient catastrophes.
Finally, at zero temperature and using a local density approximation for the initial fluid state, we compare the hydrodynamic prediction for the space-time density profile with a simulation of the full quantum model obtained from the NRG-TSA-ABACUS algorithm \cite{abacus,konik,caux_konik}, and find perfect agreement. 

\vspace{0.2cm}

\noindent {\bf\em GHD.}\quad The Hamiltonian of the repulsive LL model is
\beq\label{LL}
	H = \int \dd x\,\lt(\frc1{2m}\p_x\psi^\dag \p_x\psi + \frc c2\psi^\dag\psi^\dag\psi\psi\rt),\quad c>0
\eeq
for the complex bosonic field $\psi(x)$, where $m$ is the mass (we set $\hbar=1$ throughout the manuscript). An inhomogeneous initial state $\bra\cdots\ket$, to be specified below, is set to evolve unitarily with $H$.

Since the LL model is integrable, it admits infinitely many conservation laws $\p_t q_i + \p_x j_i=0$. This includes the gas density $q_0 = \psi^\dag\psi$, the momentum density $q_1=-\ri \psi^\dag\p_x\psi + h.c.$, and the energy density $q_2$ (the integrand in \eqref{LL}).  According to the principles of hydrodynamics, if averages of conserved densities $\bra e^{\ri H t} q_i(x)e^{-\ri H t} \ket$ and currents $\bra e^{\ri H t} j_i(x)e^{-\ri H t} \ket$ have smooth enough space-time profiles, they can be described by space-time dependent local states that have maximized entropy with respect to the conserved charges afforded by the dynamics. Eulerian hydrodynamics, which neglects viscosity effects and is valid at large scales, is formed of the ensuing macroscopic conservation laws. In integrable systems, entropy maximization leads to generalized Gibbs ensembles (GGEs) \cite{thermalization,GGEreviews} with (formal) density matrix $\rho_{\rm GGE} = e^{-\sum_i \beta_i Q_i}$, $Q_i= \int q_i (x) \dd x$. Therefore,  $\bra e^{\ri H t}\Or(x)e^{-\ri H t}\ket \approx {\rm tr}[\rho_{\rm GGE}(x,t) \Or]$, where the only space-time dependence is in $\rho_{\rm GGE}(x,t)$. The macroscopic conservation laws of generalized hydrodyamics (GHD) are the infinite number of equations for the density averages ${\tt q}_i(x,t) = {\rm tr}[\rho_{\rm GGE}(x,t) q_i]$ and the current averages ${\tt j}_i(x,t)={\rm tr}[\rho_{\rm GGE}(x,t) j_i]$:
\begin{equation}\label{conserv}
 \p_t{\tt q}_i+\p_x{\tt j}_i=0.
\end{equation}
The set of ${\tt q}_i$ fixes the GGE state, and thus can be seen as a set of fluid variables for GHD. In the manifold of GGE states, the currents ${\tt j}_i$ have a fixed functional form in terms of the densities ${\tt q}_i$: these are the equations of state, which fully determine the GHD model at hand. 

An efficient treatment of hydrodynamics requires an appropriate choice of fluid variables. Instead of the ${\tt q}_i$, the most powerful fluid variables are obtained in terms of the emerging quasi-particles of the integrable model. In the repulsive LL model, there is a single quasi-particle species. The interaction in the LL model is fully described by the two-particle scattering matrix $S(\theta-\theta')$, a function of velocity differences. The object of importance is the differential scattering phase \cite{LiebLiniger},
\beq
	\varphi(\theta)=-\ri \,\frac{\dd}{\dd \theta} \log S(\theta) = 2c/(\theta^2+c^2).
\eeq
The quasi-particle can be seen as a spinless real fermion, which is free in the Tonks-Girardeau (TG) limit $c=\infty$ (hard-core repulsion). 

States $|\theta_1,\ldots,\theta_N\ket$ are described by the velocities $\theta_k$ of the quasi-particles. Each conserved charge $Q_i$ is characterized by its one-particle eigenvalue $h_i(\theta) \propto \theta^i$, with $Q_i|\theta,\ldots,\theta_N\rangle=\sum_k h_i(\theta_k)|\theta_1,\ldots,\theta_N\rangle$. For instance, the particle number has eigenvalue $h_0(\theta) = 1$, the momentum $h_1(\theta)=p(\theta)=m\theta$, and the energy $h_2(\theta)=E(\theta)=m\theta^2/2$. In the thermodynamic limit, the eigenstates are expressed in terms of $\rho_{\rm p}(x,\theta)\dd x\dd\theta$, the number of quasi-particles in the phase-space region $[x,x+\dd x]\times [m\theta,m(\theta+\dd\theta)]$, leading to average densities ${\tt q}_i = \int \dd\theta\, \rho_{\rm p}(\theta) h_i(\theta)$. The most convenient fluid variable is the occupation function $n(\theta)=\rho_{\rm p}(\theta)/\rho_{\rm s}(\theta)$, where $\rho_{\rm s}$ is the state density, $2\pi \rho_{\rm s}(\theta)=m + \int \dd \alpha\,\varphi(\theta-\alpha)\rho_{\rm p}(\alpha)$. The density and current averages take the form \cite{ghd},
\beq\label{qjn0} 
	{\tt q}_i = m\int \frc{\dd\theta}{2\pi}\,
	n(\theta)\,h_i^{\rm dr}(\theta),\quad
	{\tt j}_i = m\int \frc{ \dd\theta\,\theta}{2\pi}\,
	n(\theta)\,h_i^{\rm dr}(\theta)
\eeq
where the dressing operation is defined by
\begin{equation}\label{dress}
 f^{\dr}(\theta)=f(\theta)+\int\frac{\dd \alpha}{2\pi}\,\varphi(\theta-\alpha)n(\alpha)f^{\dr}(\alpha).
\end{equation}
These establish a relation between the ${\tt j}_i$'s and the ${\tt q}_i$'s, and thus the equations of state.  For the LL model they were derived in \cite{ghd} by extending the theory of (generalized) TBA \cite{caux_konik,GTBA}.

It was realized in \cite{ghd,bertini1} that demanding the continuity
equations \eqref{conserv} together with the averages \eqref{qjn0} implies the continuity equation at the level of quasi-particles: 
\begin{equation}\label{nconti}
 \p_tn(\theta)+v^{\rm eff}(\theta)\p_xn(\theta)=0,
\end{equation}
where the effective velocity $v^{\rm eff}(\theta)$ is the velocity of elementary excitations \cite{vexcitations}
\beq\label{veos}
	v^{\rm eff}(\theta) = \frac{ (E')^{\dr}  (\theta) }{(p')^{\dr}(\theta) } = \frac{{\rm id}^{\rm dr}(\theta)}{1^{\rm dr}(\theta)}
\eeq
with ${\rm id}(\theta) = \theta$. These are the GHD equations in terms of quasi-particle fluid variables in the LL model. Since $v^{\rm eff}(\theta)$ depends on the fluid state through the function $n$, these are nonlinear equations for an infinity of functions of space-time (one for each velocity $\theta$).

Some intuition into Eqs. (\ref{nconti}), (\ref{veos}) can be gained by looking at the TG limit $c=\infty$. In this case, $n(\theta)$ is  the fermion occupation number at each momentum $m\theta$, at position $x$. This is the Wigner function of the state \cite{wigner} (the partial Fourier transform of the fermion-fermion correlator). The effective velocity is equal to the particle velocity, $v^{\rm eff}(\theta) = \theta$, and  (\ref{nconti}) simply reproduces the exact evolution equation for the Wigner function, a direct consequence of the Schr\"odinger equation, as exploited in \cite{bumps_free_fermions,Wigner_oscillations} (see also the Supplementary Material (SM)). The quasi-particle occupation $n(\theta)$ may thus be viewed as the generalization of the Wigner function to non-free-fermion systems, with time-evolution governed by GHD (\ref{nconti})-(\ref{veos}).

\vspace{0.2cm}

\noindent {\bf\em Zero-entropy GHD.}\quad  Natural initial conditions are ground states within inhomogeneous potentials $V(x)$,
\beq\label{HV}
	H_V = \int \dd x\,\lt(\frc1{2m} \p_x\psi^\dag \p_x \psi + \frc{c}2\psi^\dag \psi^\dag\psi\psi + V(x) \psi^\dag\psi\rt).
\eeq
With a slowly varying potential, local averages are well described by a local-density approximation (LDA) \cite{review_giamarchi}. LDA provides a GHD  initial condition, a fluid of local zero-temperature states, which at every point $x$ is the ground state of $H+ V(x)Q_0$.  In this section, we observe that GHD equations give rise to finite-dimensional hydrodynamics when one restricts to the subspace of fluid states with zero entropy such as those. An analogous observation was made previously for free fermions \cite{bumps_free_fermions} and for the Calogero-Sutherland model \cite{bumps_Calogero}.

Recall that the occupation function at zero temperature is $n_{T=0}(\theta)=\chi (\theta\in[-\theta_{\rm F},\theta_{\rm F}])$ (where $\chi$ is the indicator function) where $\theta_{\rm F}$ is the Fermi pseudo-velocity, which depends on the chemical potential.
Let us consider the space of zero-entropy occupation functions which have exactly $2k$ jumps
, characterized by $2k$ velocities $\cdots<\theta_{j-1}^+<\theta_j^-<\theta_j^+<\theta_{j+1}^-<\cdots$ bounding separate Fermi seas: $n(\theta)= \chi(\theta\in
\cup_{j=1}^k[\theta_j^-,\theta_j^+]))$. We show that under GHD evolution, any smooth fluid whose state lies in such a space at all positions $x$, stays so for short enough times. Time evolution leads to displacements of Fermi points. Thus at zero entropy, GHD is reduced to hydrodynamics with a finite number of fluid variables.

Indeed we have $\p_x n(\theta) = -\sum_{\ep=\pm}\sum_{j=1}^k \ep\p_x \theta_j^\ep\,\delta(\theta-\theta_j^\ep)$, and thus the time derivative $\p_tn(\theta)$ is supported on the finite set of velocities $\theta_j^\pm$. A solution to \eqref{nconti} is therefore provided by setting $\theta^\pm_j = \theta^\pm_j(x,t)$ with
\beq\label{khd}
	\p_t \theta^\pm_j + v^{\rm eff}_{\{\theta\}}(\theta^\pm_j) \p_x\theta^\pm_j = 0.
\eeq
We expect the solution to \eqref{nconti} in the space of smooth fluid space-time functions to be unique, based on such rigorous results in related classical gases \cite{dobrods}. Thus it is given by solving \eqref{khd} as long as no shock develops. Here, more explicitly, the effective velocity is $v^{\rm eff}_{\{\theta\}}(\alpha) = {\rm id}^{\rm dr}_{\{\theta\}}(\alpha)/1^{\rm dr}_{\{\theta\}}(\alpha)$ with the dressing operation $f^{\rm dr}_{\{\theta\}}(\alpha) = f(\alpha) + \sum_{j=1}^k
	\int_{\theta_j^-}^{\theta_j^+}\dd \gamma\,
	\varphi(\alpha-\gamma)f^{\rm dr}_{\{\theta\}}(\gamma)$. The resulting equations \eqref{khd} will be referred to as $2k$-hydrodynamics ($2k$HD). 

\begin{figure}[th]
	a.\includegraphics[width=0.47\textwidth]{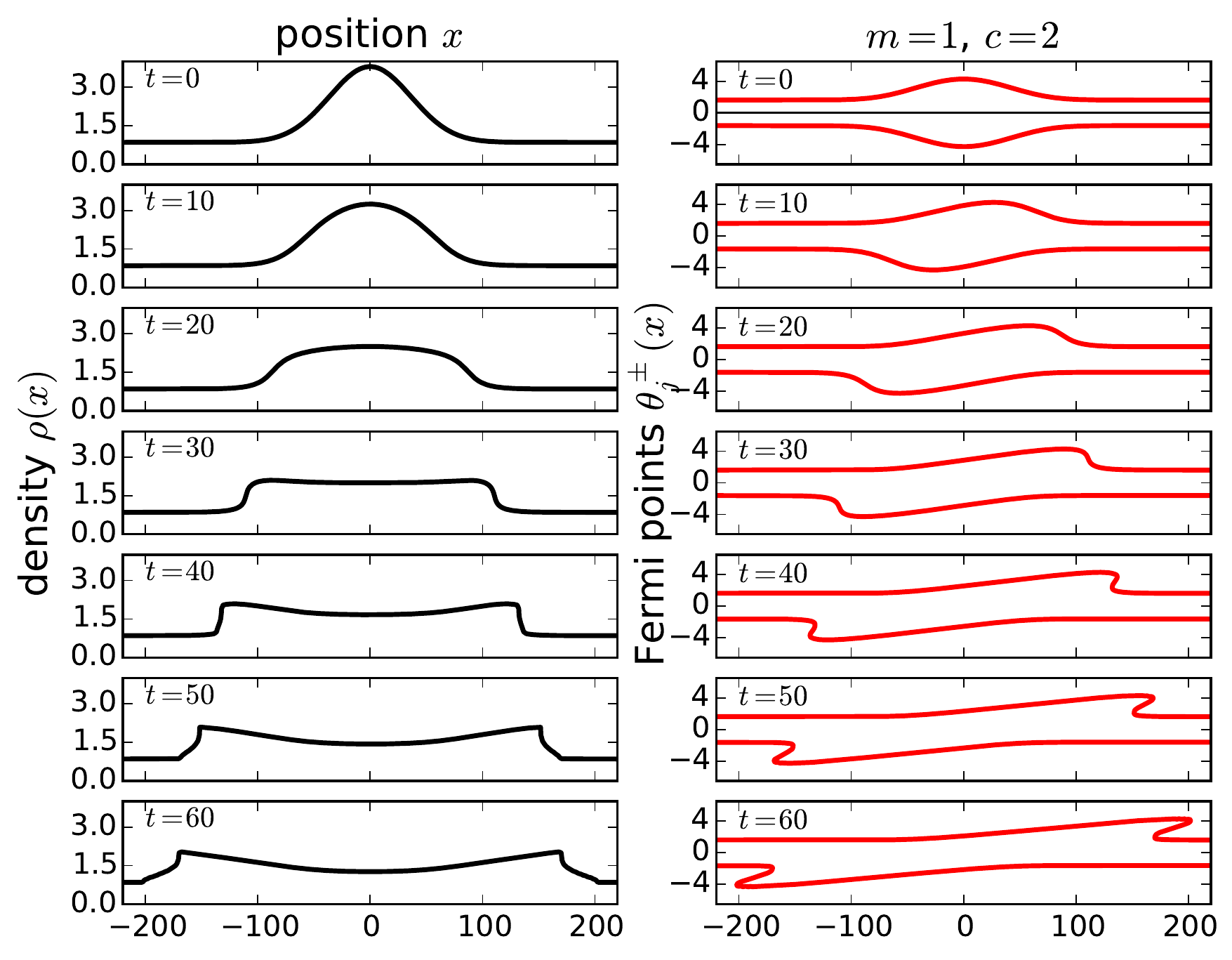}
    b.\includegraphics[width=0.42\textwidth]{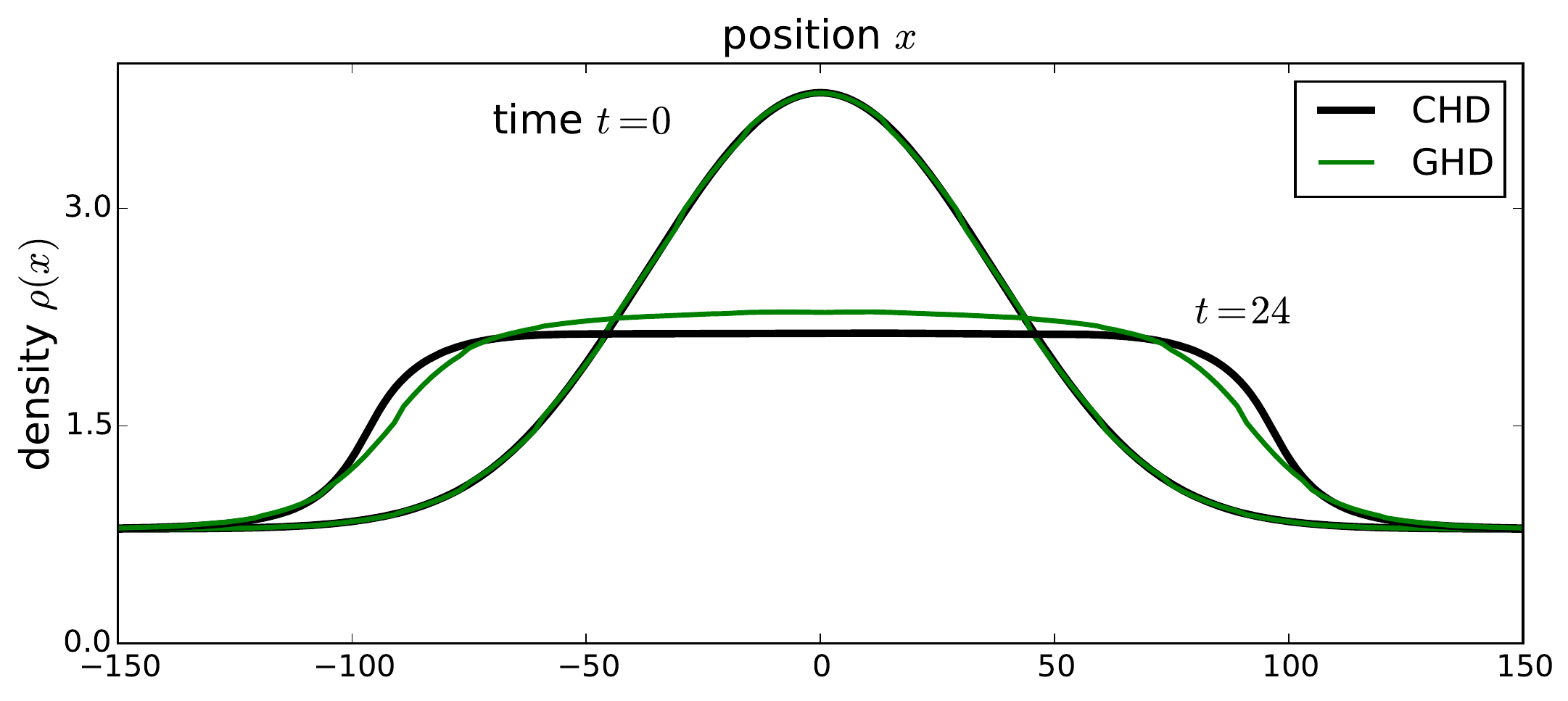}
    \caption{(a) Left: density profile of LL gas suddenly released from a gaussian potential $V(x) = -5 e^{-(\frac{x}{50})^2}-1$. Right: Corresponding Fermi points $\theta_j^{\pm} (x)$. Initially, there are only $k=2$ Fermi points, but after the shock at $t \simeq 37$, there is a region where the red curve is multi-valued, corresponding to $k=4$ Fermi points.
    (b) Same setup at finite temperature: the initial state is obtained from LDA at temperature $T=1$. After finite time, CHD quantitatively differs from GHD. Moreover, at a later time $t \simeq 35$, CHD has a shock (see the SM); in contrast, GHD has no shock.}
    \label{fig:shock_2HD_4HD}
\end{figure}

Eqs \eqref{khd} are Euler-type hydrodynamic equations for a fluid with finitely-many components. Finite-component fluids are expected to develop shocks.
Therefore, Eqs \eqref{khd} are expected to hold {\em only for finite times}. However, contrary to true conventional finite-component fluids, where viscosity effects, present beyond the Euler scale, dominate and produce entropy at shocks, the presence of infinitely-many conservation laws forbids sustained entropy production in GHD. Any shock instantaneously dissolves into the higher-dimensional solution space of GHD. More precisely, $2k$HD solutions become multivalued at the time of the appearance of the shock, but here this multivaluedness is physically meaningful, representing a higher number of Fermi seas. Thus shocks in $2k$HD resolve by increasing the number of Fermi seas, passing to $(2k+2)$HD. We exemplify this in Fig. \ref{fig:shock_2HD_4HD}a, where after $t_{shock}=37$ we begin to simulate, effectively, 4HD equations. This has previously been observed in free fermion models \cite{bumps_free_fermions}, thanks to an analysis based on the Wigner function; here it is generalized to the fully interacting LL model. We have also confirmed this shock dissolution mechanism of GHD in a nontrivial classical gas with the same hydrodynamic equations as those of the LL model, see the SM.


\vspace{0.2cm}
\noindent {\bf\em GHD and conventional hydrodynamics (CHD).}\quad Starting with a smooth fluid of local zero-temperature states, GHD reduces to 2HD,
where every local fluid cell is the Galilean boost of a zero-temperature state. As a consequence, 2HD is in fact equivalent to the conventional hydrodynamics (CHD) of Galilean fluids,
\beq
	\p_t\rho+\p_x(v\rho )=0,\quad \p_tv+v\p_xv=-\frac{1}{m\rho}\p_x \mathcal{P}\label{chd}
\eeq
where $\rho={\tt q}_0$ is the fluid density and $\mathcal{P}$ is the pressure (for a proof of the equivalence, see the SM). The first equation is conservation of mass, the second, of momentum. The pressure $\mathcal{P}=\mathcal{P}(\rho)$ gives the equations of state of the fluid, and here equals the momentum current ${\tt j}_1$ in the zero-temperature state with density $\rho$. The explicit equations of state are obtained from \eqref{qjn0} and \eqref{dress} (see SM).

CHD has been used as an important tool in analyzing the dynamics of 1d Bose gases \cite{chdpapers,damski,pv14}. 
It has sometimes been presented as a consequence of the Gross-Pitaevskii equation \cite{review_giamarchi}---itself valid only in the limit of small interaction strength $c$---, and it was never quite clear what exactly the range of validity of CHD was in the full interaction range of the LL model. Our analysis clearly shows that CHD is valid ---in the sense that it coincides with GHD--- {\it only at zero temperature}, and {\it before the first shock}. We conclude that, {\it in any other situation, CHD is not applicable and leads to quantitatively wrong results}. To illustrate this, a comparison of CHD at finite temperature and GHD is shown in Fig.\ref{fig:shock_2HD_4HD}b; the initial state is the same in both cases
(obtained from LDA at finite temperature), but one sees that the density profiles differ significantly at finite time; moreover, CHD has solutions up to a finite shock time, while GHD has no shocks and has solutions at  arbitrarily long times.

\vspace{0.2cm}

\begin{figure}[ht]
	\begin{center}
    	\includegraphics[width=8.0cm]{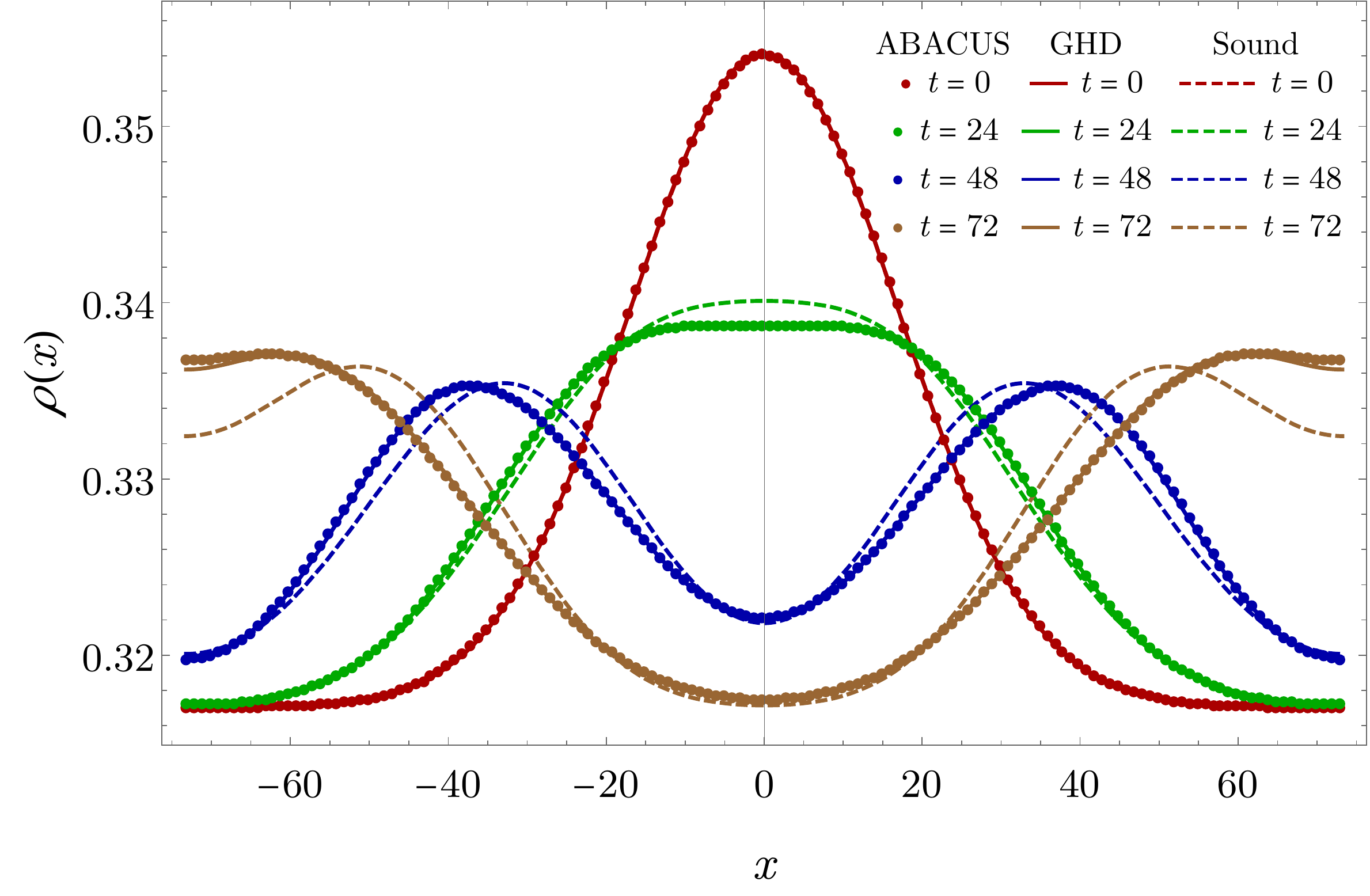}
        \caption{The density profile under the evolution with $2m=1$, $c=1$, $U=0.03$, $a=1/576$, and a choice of $\mu_\infty$ such that
there are $N = 48$ particles in the ground state. We use periodic boundary conditions. Points show NRG-TSA-ABACUS data, full line the GHD simulation, and dotted line a linear sound wave approximation.}
       \label{f1}
    \end{center}
\end{figure}

\noindent {\bf\em Comparison with microscopic simulation of the LL model.}\quad 
We consider evolution from the ground state of \eqref{HV} with a background chemical potential $\mu_\infty$ perturbed by a Gaussian, $V(x) = -\mu_\infty - Ue^{-ax^2}$. The initial density profile accumulates around $x=0$, and is asymptotically nonzero. Two procedures are compared: (1) the ground state is exactly evaluated using the NRG-TSA-ABACUS algorithm \cite{konik,caux_konik} (see the SM for details), and then evolved unitarily; and (2) the ground state is approximated using LDA, and this initial fluid state is evolved using 2HD (see the SM for a review of standard conditions for the hydrodynamic regime, which are fulfilled by the choice of parameters below). Fig. \ref{f1} provides the result for a choice of parameters corresponding to the local dimensionless coupling $\gamma(x)=mc/\rho(x)$ of the order of 1,
thus the system is in an intermediate regime with nontrivial interactions being important. We observe that GHD is in excellent agreement with NRG-TSA-ABACUS numerics at almost all times except near the right and left boundaries at $t=72$, and provides a substantially better approximation than
linear sound waves (see the SM). It can also be seen that, since $v^{\rm eff}_{\{\theta\}}(\theta^+)$ is always greater than the background Fermi velocity  corresponding to $\mu_{\infty}$, the propagation speed of 2HD is larger than that of the sound wave. It is remarkable that the complex (zero-temperature) dynamics of the LL gas is exactly described by 2HD, a simple set of differential equations.


At large times, discrepancies, though very tiny in the above graph,  are expected to emerge. One reason is that, as explained above, shocks attempt to form, and as variations become larger, conditions for the hydrodynamic regime break down. Higher-derivative effects, such as viscosity, become more important.
Recent observations in the related hard rod gas \cite{ds} suggest however that such higher-derivative effects play only a small role. Another cause for late-time discrepancy is that LDA is not exact. As higher-order charges are more sensitive to the large-scale variations of the potential, LDA gives an extremely good approximation to the particle density, but describes poorly densities of higher-order charges $Q_i$. As time passes, the effects of the latter under the full GHD evolution eventually breaks 2HD. An analysis of the Wigner function $n(\theta)$ at the free-fermion point $c=\infty$, where GHD is exact (no viscosity is neglected), gives further insight (see the SM).

\vspace{0.2cm}

\noindent {\bf\em Conclusion.}\quad We showed that widely used conventional hydrodynamics of the Lieb-Liniger model correctly describes interacting Bose gases, but that this holds only at zero temperatures and for finite times. We provided exact, simple hydrodynamic equations valid beyond gradient catastrophes, where no shocks are sustained. These are zero-entropy reductions of GHD, which are finite-component fluid equations easily solvable on a laptop. This suggests that fluids of integrable models avoid entropy production thanks to the large space of fluid states. We provided compelling evidence for the emergence of GHD in the LL gas in the limit of slow variations of the density profile. This provides a crucial dynamical extension of LDA that is valid beyond previously existing frameworks.
As a future direction, it would be very interesting to apply our method to more experimentally relevant situations such as ``Quantum Newton's cradle"-type protocol \cite{qnc}, known to be beyond the reach of CHD \cite{pv14}.

\vspace{0.2cm}

\noindent {\bf\em Acknowledgements:} We are extremely grateful to J.-S. Caux for access to the ABACUS software package without which the Lieb-Liniger simulations reported here would not have been possible. We thank B. Bertini, M. Fagotti, and H. Spohn for stimulating discussions, and the anonymous referees for useful comments. TY is grateful for the support from the Takenaka Scholarship Foundation and the hospitality at the Tokyo Institute of Technology. BD and JD thank the IESC Carg\`ese for hospitality.  RMK's research effort here was supported by the U.S. Department of Energy, Office of Basic Energy Sciences, under Contract No. DE- AC02-98CH10886.


\newpage

\pagebreak
\begin{center}
\textbf{\large Supplementary material}
\end{center}

\section{Conventional hydrodynamics and sound wave}\label{2HD=CHD}

\subsection{2HD = CHD}

In the set of zero-entropy states, GHD boils down to 2HD, the hydrodynamics of two conserved quantities described by 
Eq. (9) in the main text, with $k=2$. Here we show that 2HD is equivalent to the conventional hydrodynamics (CHD) of Galilean fluids defined by Eqs. (10) in the main text.

First, Eq. (9) simplifies thanks to Galilean invariance. Let
\beq
	v_{\rm F}(\theta) = v^{\rm eff}_{\{-\theta,\theta\}}(\theta)
\eeq
be the equilibrium Fermi velocity, with Fermi pseudo-velocity $\theta$.  Let $\Lambda = (\theta^+-\theta^-)/2$ and $\eta = (\theta^++\theta^-)/2$. Shifting the integration variable by $\eta$ in the dressing operation, we obtain $f^{\rm dr}_{\{\theta^-,\theta^+\}} = \big(f\circ \iota_{\eta}\big)^{\rm dr}_{\{-\Lambda,\Lambda\}}\circ\iota_{-\eta}$ where $\iota_\eta(\theta) = \theta+\eta$ is the shift function. Using ${\rm id}\circ\iota_\eta = {\rm id}+\eta$ and $1\circ\iota_\eta=1$, we find $v^{\rm eff}_{\theta^-,\theta^+}(\alpha) = v_{\rm F}(\alpha-\eta)+\eta$, and with $v_{\rm F}(-\alpha) = -v_{\rm F}(\alpha)$ it follows that $v^{\rm eff}_{\{\theta^-,\theta^+\}}(\theta^{\pm})=\pm v_{\rm F}(\Lambda)+\eta$. This leads to
\beq\label{t0eq2}
 \p_t\theta^\pm+\Big((\theta^++\theta^-)/2 \pm v_{\rm
 F}\big((\theta^+-\theta^-)/2\big)\Big)\p_x\theta^\pm=0.
\eeq
That is, the hydrodynamics is completely determined by the functional form of the equilibrium Fermi velocity. Notice that the propagation velocity $\pm v_{\rm F}(\Lambda)+\eta$ of $\theta^\pm$ equals $\pm v_{\rm F}(\Lambda)+v$, where $v = {\tt j}_0/{\tt q}_0 = {\tt q}_1/{\tt q}_0=\eta$ is the fluid velocity. This is a simple consequence of the Galilean velocity-addition formula. We note that this 2HD equation \eqref{t0eq2} was in fact already derived in \cite{bump_ll_A}, although the authors did not provide the interpretation that $\theta^\pm$ are``dynamical" Fermi pseudo-velocities.


Next, we observe that the equation of state of CHD -- the relation between the pressure and the density in Eq. (10) in the main text -- is obtained as follows. The pressure is given in two equivalent ways
\beq\label{P1}
	\mathcal{P}(\rho) = m^2\int_{-\Lambda_\rho}^{\Lambda_\rho} \frac{\dd \theta\,\theta}{2\pi}
	U(\theta)=-m^2\int_{\Lambda_\rho}^{\Lambda_\rho}\frac{\dd \theta}{2\pi}\mathcal{E}(\theta)
\eeq
where the function $U(\theta)$ and $\mathcal{E}(\theta)$ solve
\beq
U(\theta) = \theta + \int_{-\Lambda_\rho}^{\Lambda_\rho} \frac{\dd\alpha}{2\pi}\varphi(\theta-\alpha)U(\alpha),
\eeq
\begin{equation}
\mathcal{E}(\theta) = \frc{\theta^2}{2}-\frac{\mu(\Lambda_\rho)}{m} + \int_{-\Lambda_\rho}^{\Lambda_\rho} \frac{\dd\alpha}{2\pi}\varphi(\theta-\alpha)\mathcal{E}(\alpha),
\end{equation}
where $\mathcal{E}(\theta)$ is defined so that it satisfies
$\mathcal{E}(\Lambda_\rho)=0$ from which $\mu(\Lambda_\rho)$ is also fixed. The Fermi pseudo-velocity $\Lambda_\rho$ is determined by
\beq\label{P2}
	\rho = m\int_{-\Lambda_\rho}^{\Lambda_\rho} \frac{\dd \theta}{2\pi}
	V(\theta)
\eeq
where the function $V(\theta)$ solves
\beq
V(\theta) = 1 + \int_{-\Lambda_\rho}^{\Lambda_\rho} \frac{\dd\alpha}{2\pi}\varphi(\theta-\alpha)V(\alpha).
\eeq
Observe that $U(\theta), V(\theta)/2\pi,$ and $\mathcal{E}(\theta)$ are
nothing but the dressed momentum, the density of state, and the
pseudo-energy, respectively, at $T=0$ with the Fermi pseudo-velocity $\Lambda_\rho$.

Using these expression, we can explicitly demonstrate that the
CHD equations are recast into Eqs. \eqref{t0eq2}. 
The derivative of
the density $\rho(x,t)=\rho(\Lambda(x,t))$ with respect
to $i$ ($i=x,t$) can be written in the following way
\begin{equation}
 \p_i\rho(\Lambda)=\p_i\Lambda\frac{\p \mu(\Lambda)}{\p
 \Lambda}\frac{\p \rho(\Lambda)}{\p\mu(\Lambda)}=\p_i\Lambda mv_{\F}(\Lambda)\kappa(\Lambda),
\end{equation}
where $\kappa(\Lambda):=\p\rho(\Lambda)/\p\mu(\Lambda)$ is the
compressibility. Here we also used $\p\mu(\Lambda)/\p\Lambda=mv_{\rm
F}(\Lambda)=mU(\Lambda)/V(\Lambda)$ where the first equality can be
shown by differentiating $\mathcal{E}(\Lambda)=0$ with respect to
$\Lambda$. Using an identity $\kappa(\Lambda)=V^2(\Lambda)/(\pi v_{\rm
F}(\Lambda))$ \cite{KBI_A}, we obtain
\begin{equation}
 \p_i\rho(\Lambda)=\frac{\p_i\Lambda mV^2(\Lambda)}{\pi}.
\end{equation}
The density itself also bears a similar expression
$\rho(\Lambda)=mv_{\F}(\Lambda)\rho_{\rm s}^2(\Lambda)/\pi$ in terms of
$v_{\rm F}(\Lambda)$ and $V(\Lambda)$ \cite{KBI_A}. An analogous manipulation on derivatives of the pressure
$\mathcal{P}(\Lambda)=\mathcal{P}(\rho(\Lambda))$ is done using a thermodynamic relation $\p \mathcal{P}/\p \mu=\rho$,
which readily follows from \eqref{P1}, yielding
\begin{equation}
 \p_i\mathcal{P}(\Lambda)=\p_i\Lambda mv_{\F}(\Lambda)\rho(\Lambda).
\end{equation}
 Necessary ingredients are now ready; combining them and $v={\tt j}_0/{\tt q}_0=\eta$, the continuity equation and the Euler equation in 
 \eqref{chd} become, respectively,
\begin{equation}
 \begin{aligned}
  \p_t\Lambda+\eta\p_x\Lambda+v_{\F}(\Lambda)\p_x\eta&=0,\\
 \p_t\eta+\eta\p_x\eta+v_{\F}(\Lambda)\p_x\Lambda&=0,
\end{aligned}
\end{equation}
which are clearly equivalent to \eqref{t0eq2}.

\subsection{Sound wave}
When the fluid of the LL gas is in the linear-response regime, we can describe the fluid as a sound wave: expanding $\theta^\pm(x,t)=\pm\theta_{\rm F}+\delta\theta^\pm(x,t)$, where $\theta_{\rm F}$ is the Fermi-pseudo velocity of the unperturbed LL gas, the 2HD equation \eqref{t0eq2} becomes
\begin{equation}
\partial_t\delta\theta^\pm(x,t)\pm v_{\rm F}(\theta_{\rm F})\p_x\delta\theta^\pm(x,t)=0.
\end{equation}
This is the sound wave propagating with the Fermi velocity $v_{\rm F}(\theta_{\rm F})$, and describes the dynamics of the linear Luttinger liquid. Note that, in the LL gas, the Fermi velocity is same as the sound velocity $v_{\rm s}=\sqrt{\p \mathcal{P}/\p(m\rho)}$, which is readily confirmed.

\subsection{Conventional hydrodynamics at finite temperature}

Here we explain how to generalize the CHD Eqs. (10) in the main text to the finite temperature case, needed to produce the data in Fig. (1.b) in the main text, see also Fig. \ref{fig:finite_T_appendix} below.

To do CHD at finite temperature, one needs to keep track of three conserved quantities: the number of particles, the momentum, and the energy. These correspond to the three densities ${\tt q}_0$, ${\tt q}_1$ and ${\tt q_2}$ respectively. In principle, one can write three continuity equations for these densities, which can be recast in the more conventional form of Euler hydrodynamics
\begin{equation}
	\begin{array}{rcl}
    	\partial_t n + \partial_x (n u) &=& 0 \\
       ( \partial_t + u \partial_x ) u &=& - \frac{1}{\rho} \partial_x P \\
       (\partial_t + u \partial_x) \tau + \frac{2}{3} (\partial_x u ) \tau &=& 0
    \end{array}
\end{equation}
where $n$ is the particle density, $u$ is the local velocity of the fluid, $\tau$ is the kinetic energy, and $P$ is the pressure.

We find it more convenient, however, to work directly with the continuity equations
\begin{equation}
	\partial_t {\tt q}_i +  \partial_x {\tt j}_i \, = \, 0, \qquad i=0,1,2, 
\end{equation}
and with the Lagrange multipliers $\beta_0$, $\beta_1$ and $\beta_2$, such that the reduced density matrix at a point $(x,t)$ is approximated by the Generalized Gibbs Ensemble with three charges, $e^{-\beta_0 Q_0 - \beta_1 Q_1  -\beta_2 Q_2 }$ where $Q_0 = N$ is the number of particles, $Q_1 = P$ is the momentum, $Q_2 = H$ is the Hamiltonian, and
$\beta_2$ is the inverse temperature while $-\beta_0/\beta_2$ is the chemical potential. 
The occupation function $n(\theta)$ in this Generalized Gibbs Ensemble is obtained by solving the (Generalized) Thermodynamic Bethe Ansatz equations,
\begin{eqnarray}
\nonumber \frac{\varepsilon(\theta)}{T} &=& \sum_{i=0,1,2} \beta_i q_i(\theta) - \int \frac{d\theta'}{2\pi} \varphi(\theta-\theta') \log [1+e^{-\frac{\varepsilon(\theta')}{T}}] \\
    n(\theta) &=& \frac{1}{ 1+ e^{\frac{\varepsilon(\theta)}{T}}} ,
\end{eqnarray}
where $q_0(\theta)=1$, $q_1(\theta)= m \theta$, $q_2(\theta) = m\theta^2/2$.
Then, using the results of Ref.  \cite{doyon_spohn_A}, one gets
\begin{eqnarray*}
	\frac{\partial {\tt q}_j}{\partial \beta_i} &=&  - \int \frac{d \theta}{2\pi} \rho_{\rm p}(\theta) (1-n(\theta)) h_i^{\rm dr} (\theta) h_j^{\rm dr} \\
    \frac{\partial {\tt j}_j}{\partial \beta_i} &=&  - \int \frac{d \theta}{2\pi} \rho_{\rm p}(\theta) v^{\rm eff}(\theta) (1-n(\theta)) h_i^{\rm dr} (\theta) h_j^{\rm dr} .
\end{eqnarray*}
The CHD equations at finite temperature can then be written as
\begin{equation}
	\sum_{j=0}^2 \left[ C_{i j}\partial_t \beta_j (x,t) + B_{i j}\partial_x \beta_j (x,t) \right] = 0,
\end{equation}
where
\begin{eqnarray*}
	C_{ij} &=&  \int \frac{d \theta}{2\pi} \rho_{\rm p}(\theta) (1-n(\theta)) h_i^{\rm dr} (\theta) h_j^{\rm dr} \\
   B_{ij} &=& \int \frac{d \theta}{2\pi} \rho_{\rm p}(\theta) v^{\rm eff}(\theta) (1-n(\theta)) h_i^{\rm dr} (\theta) h_j^{\rm dr} .
\end{eqnarray*}
These are the equations we solve to produce the data for CHD at finite temperature in Fig. (1.b) in the main text and in Fig. \ref{fig:finite_T_appendix}.

\begin{figure}[ht]
	\begin{center}
    	\includegraphics[width=8.cm]{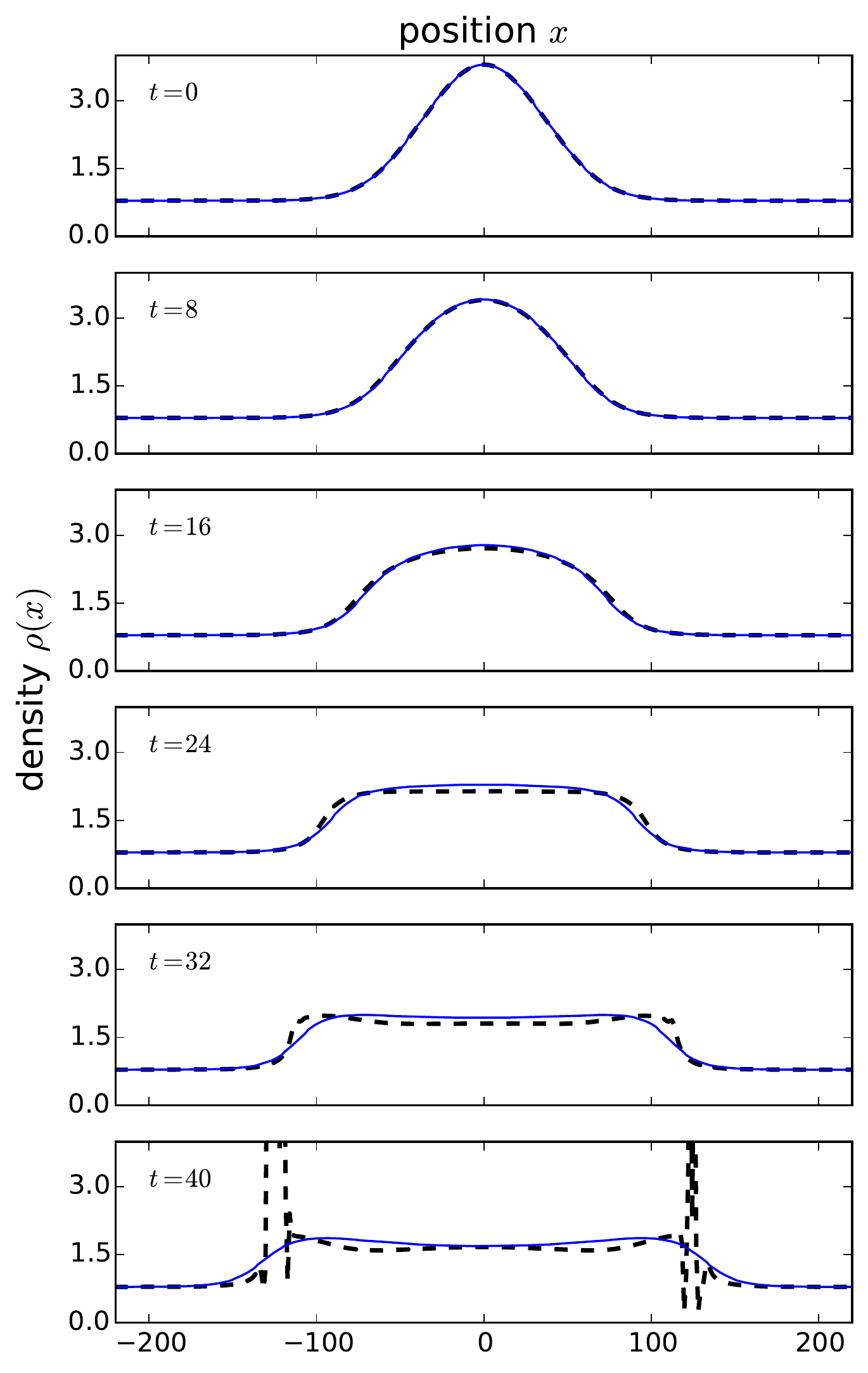}
        \caption{Evolution of the density profile at temperature $T=1$, with parameters $m=1$, $c=2$ and an initial potential $V(x) = -5 e^{-(\frac{x}{50})^2}-1$. The solution from CHD (black, dashed) is compared to the GHD solution (blue) obtained from the molecular simulation of the classical flea gas. We see that, already at $t=24$, CHD differs significantly from GHD. We see CHD has a shock at $t \simeq 35$, and loses its validity after the shock. GHD has no shocks.}
       \label{fig:finite_T_appendix}
    \end{center}
\end{figure}

\section{The free fermion case}

Hydrodynamics of 1d free fermions has been studied thoroughly in the past decade \cite{bumps_free_fermions_A,Wigner_oscillations_A}, with the conclusion that the leading role is played by the Wigner function. For the convenience of the reader, we now discuss the aspects of the free fermion case that shed light on GHD, in relation with the discussion in the main text; for full details about the free fermion case, the reader is referred to the original references  \cite{bumps_free_fermions_A,Wigner_oscillations_A}.

\subsection{GHD for free fermions, and the role of the Wigner function}

In the Tonks-Girardeau limit $c \rightarrow \infty$, the bosons are impenetrable. Impenetrable bosons can be mapped to non-interacting fermions through the Jordan-Wigner mapping,
\begin{equation*}
	\Psi_{\rm F}^\dagger (x) \,: = \, e^{i \pi \int_{y< x} \psi^\dagger(y) \psi(y) dy}  \, \psi^\dagger(x) 
\end{equation*}
and a similar relation for the annihilation operator $\Psi_{\rm F}$, such that they satisfy the canonical anticommutation relations $\{ \Psi^\dagger(x) , \Psi(x') \} = \delta (x-x')$ and $\{ \Psi(x) , \Psi(x') \} = \{ \Psi^\dagger(x) , \Psi^\dagger(x') \} = 0$. 
In terms of these operators, the Hamiltonian of the Lieb-Liniger model with $c \rightarrow +\infty$, or Tonks-Girardeau gas, is simply
\begin{equation}
	H_{\rm TG} \, = \, \int dx  \; \Psi_{\rm F}^\dagger (x) \left(   -\frac{1}{2m}  \partial^2_x    \right) \Psi_{\rm F} (x) .  
\end{equation}
What is crucial here is that the hamiltonian is quadratic, so the particles created/annihilated by $\Psi^\dagger_{\rm F}$/$\Psi_{\rm F}$ propagate
freely, and interact with other particles only through the Pauli principle---as they have fermionic statistics, unlike the original bosonic particles which entered
the definition of the model---. There is no interaction energy, and this is of course a dramatic simplification. Thanks to translation invariance $H_{\rm TG}$ is diagonal in Fourier space,
\begin{equation}
	\label{eq:TGham}
	H_{\rm TG} \,  = \, \int \frac{dk}{2\pi}    \frac{k^2}{2m}   \Psi_{\rm F}^\dagger (k)  \Psi_{\rm F} (k)
\end{equation}
where $\Psi_{\rm F}^\dagger (k) \, := \, \int  e^{i k x}  \, \Psi_{\rm F}^\dagger (x) \, dx$.

Because the Hamiltonian is quadratic and diagonal in $k$-space, we see that for any function $h(.)$, one can construct a conserved charge
\begin{equation}
	Q[h] \, := \, \int \frac{dk}{2\pi}\, h(k) \Psi^\dagger_{\rm F} (k) \Psi_{\rm F} (k).
\end{equation}
This provides a (continuous) family of commuting conserved charges,
\begin{equation*}
	\left[H, Q[h] \right]  \, = \, [Q[h] , Q[f] ] \, = \, 0 .
\end{equation*}
The associated charge density is
\begin{equation}
	q[h](x) \, = \, \int dy \, h(y) \, \Psi^\dagger (x+\frac{y}{2}) \Psi(x-\frac{y}{2})
\end{equation}
where $h(y) := \int \frac{dk}{2\pi} e^{i k y} h(k)$ is the Fourier transform of $h(k)$. For an analytic function $h(k)$, the Fourier transform decays exponentially, so the charge density $q[h](x)$ is local in the sense that it has exponentially decaying tails. 

If we allow ourselves to take $h(k) = \delta_{k_0}(k) \equiv \delta(k-k_0)$, then we obtain a charge density with tails that do not decay, but apart from that, the conserved charge $Q[\delta_{k_0}] = \frac{1}{2\pi} \Psi_{\rm F} (k_0) \Psi^\dagger_{\rm F} (k_0)$ is a perfectly decent conserved quantity. Of course, we see that it is nothing but the mode occupation at momentum $k_0$. The associated charge density reads [from now we drop the subscript '$0$' in '$k_0$']
\begin{equation}
	  q[\delta_k](x) \, =\, \frac{1}{2\pi} \int  e^{i k y}  \Psi^\dagger_{\rm F} (x + \frac{y}{2}) \Psi_{\rm F} ( x - \frac{y}{2}) dy \, .
\end{equation}

The charge density $q[\delta_{k}]$ obeys the following evolution equation, with $v(k) = k/m$,
\begin{equation}
	\label{eq:op_exact}
	\partial_t q[\delta_k](x) \, := \,  \frac{1}{i} \left[H ,\, q[\delta_k](x\right)] \, = \,  - v(k) \, \partial_x  q[\delta_k](x) \, ,
\end{equation}
which follows from a straightforward calculation using the Hamiltonian (\ref{eq:TGham}). Thus, we can identify the current operator associated to the density $q[\delta_k](x)$ as
\begin{equation}
	\label{eq:current_TG}
	j[\delta_k] (x) = v(k) q[\delta_k] (x) ,
\end{equation}
such that the continuity equation $\partial_t q[\delta_k] + \partial_x j[\delta_k] = 0$ holds. Notice also that, by linearity, this gives a simple closed formula for the current associated to any charge density $q[h]$,
\begin{equation*}
	j[h] = \int \frac{dk}{2\pi} \frac{k}{m} h(k) q[\delta_k] .
\end{equation*}
We stress that Eq. (\ref{eq:op_exact})-(\ref{eq:current_TG}) is an exact formula that holds for the operators themselves; we have not taken any expectation values yet. It is also important to emphasize that this particularly
simple form of the evolution equation is possible only because the problem is galilean invariant. If we were dealing with a lattice
gas, the velocity would not simply be $v(k) = k/m$, and there would be additional terms generated by the bracket $[H,.]$ in Eq. (\ref{eq:op_exact}).

In order to connect the simple
exact result (\ref{eq:current_TG}) for currents in the Tonks-Girardeau gas to the more general discussion of hydrodynamics in the main text, we need to turn to expectation values. Namely,
we take expectation values $\left< . \right>$ in some initial state $\left| \psi_0 \right>$ (or some initial density matrix), such that the exact relation (\ref{eq:op_exact})-(\ref{eq:current_TG}) becomes
\begin{equation}
	\partial_t {\tt q}[\delta_k] + \partial_x {\tt j}[\delta_k]  \, = \, 0 \qquad {\rm with} \quad \left\{   \begin{array}{lll}   {\tt q}[\delta_k] &:=&  \left< q[\delta_k] \right>\\
	 {\tt j}[\delta_k] &:=&  v(k) \left< q[\delta_k] \right>.
	 \end{array}
	 \right.
\end{equation}
In the language of the Thermodynamic Bethe Ansatz used in the main text, the quantity ${\rm q}[\delta_k]$ is nothing but the {\it density of particles} $\rho_{\rm p}$. In the main text, we chose to work with the occupation number $n(k) = \rho_{\rm p}(k) / \rho_{\rm s} (k)$, where $\rho_{\rm s}(k)$ is the state density. But in the Tonks-Girardeau limit, the latter is just a numerical constant, $\rho_{\rm s}(k) = 1/(2\pi)$, so we end up with
\begin{equation}
	n (x,k) =   \int dy e^{i k y} \left< \Psi_{\rm F}^\dagger(x+y/2)   \Psi_{\rm F}(x-y/2) \right> .
\end{equation}
The function $n(x,k)$ is an extremely well-known object in quantum mechanics: it is the {\it Wigner function}.
Roughly speaking, the Wigner function represents the probability of finding a particle at position $(x,k)$ in phase space. It is an extremely useful tool because it allows to get a simple (semi-classical) picture of a quantum particle, or, as we are doing here, of a large collection of non-interacting particles.

In conclusion, in the Tonks-Girardeau limit, the GHD equations are nothing but the well-known evolution equation for the Wigner function,
\begin{equation}
	\label{eq:evolution_W}
	\partial_t n(x,k,t) + v(k) \partial_x n (x,k,t) \, = \, 0 ,
\end{equation}
which can be solved trivially in terms of the Wigner function of the initial state at $t=0$,
\begin{equation*}
	n(x,k,t)	\, = \,  n(x-v(k)t,  k ,0) .
\end{equation*}
We are looking at this well-known equation in a slightly non-standard way though: for each $k$ we view $n(x,k)$, or $\rho_{\rm p}(x,k)$, as the expectation value of a single charge density $q[\delta_k]$ at position $x$. In that sense, Eq. (\ref{eq:evolution_W}) is an infinite set of continuity equations, one for each mode $k$.

\subsection{Initial Wigner function: deviation from LDA}
This section is adapted from Bettelheim and Wiegmann \cite{Wigner_oscillations_A}.

In the main text, we focus on an initial state that is the ground state of the Hamiltonian with an external potential
\begin{equation}
	\label{eq:TGham_V}
	H_{\rm TG} \,  = \, \int dx      \Psi_{\rm F}^\dagger (x) \left( - \frac{1}{2m} \partial_x^2 + V(x) \right)  \Psi_{\rm F} (x) .
\end{equation}
It is the Wigner function of this ground state that enters the formalism of GHD. In the main text, it is obtained from LDA: in the limit where the variation of $V(x)$ is small at the interparticle distance, the system is locally translation invariant, so the Wigner function should be well approximated by the one of a Fermi sea with position-dependent Fermi momentum $k_{\rm F}(x) = \sqrt{- 2 m V(x)}$,
\begin{eqnarray*}
	n_{\rm LDA} (x,k)&=& \int \frac{dq}{2\pi} e^{- i q x} \left< \Psi_{\rm F}^\dagger(k+q/2) \Psi_{\rm F}(k-q/2) \right>_{\rm F. \, sea} \\
    &=& \Theta( k_{\rm F}(x) - k )  - \Theta( -k_{\rm F}(x) - k ) ,
\end{eqnarray*}
where $\Theta(.)$ is the Heaviside function. Thus, the initial occupation number $n(x,k)$ obtained from LDA is indeed of the form discussed in the main text, $n_{\rm LDA}(x,k) = 1$ if $k^-(x)<k < k^+ (x)$, and $n_{\rm LDA}(x,k) = 0$ otherwise, for $k^+ = - k^- = k_{\rm F}$. The evolution from such an initial state is then governed by 2HD until a shock appears, which dissolves into 4HD, which itself dissolves into 6HD, and so on, as discussed in the main text.

\begin{figure}[ht]
	\begin{center}
    	\includegraphics[width=0.4\textwidth]{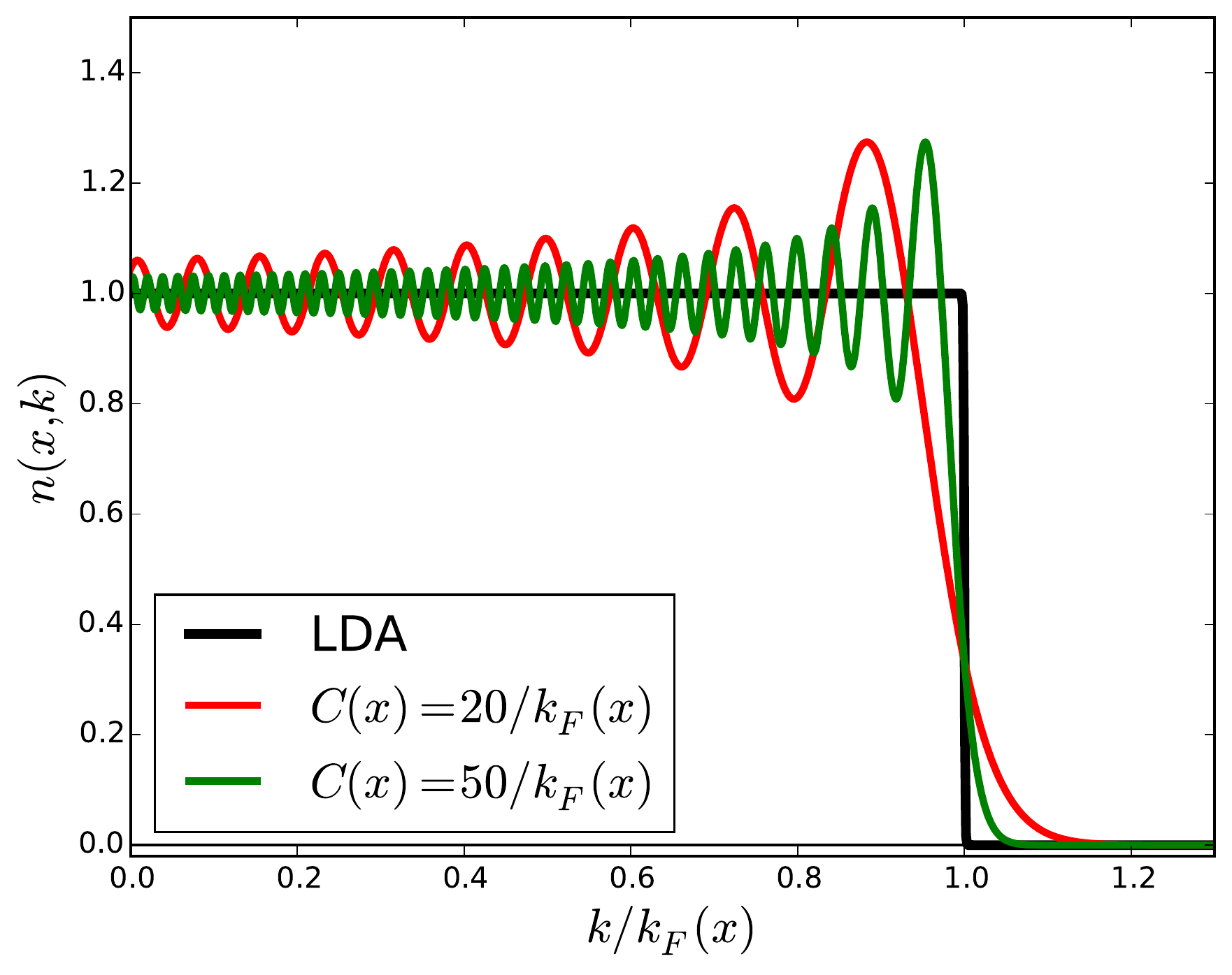} \vspace{-0.3cm}
        \caption{Wigner function $n(x,k)$, plotted as a function of $k$ at fixed $x$. The result from LDA is plotted in black. The true Wigner function, however, is oscillating around $k_F(x)$ \cite{Wigner_oscillations_A}, with a period set by the length scale $C(x) := (k_{\rm F}''(x)/8)^{-1/3}$, as illustrated by the red and green curve, given by Eq. (\ref{eq:deviation_LDA}).}
       \label{fig:Wigner_osc}
    \end{center}
\end{figure}

An important question that needs to be addressed is the accuracy of the approximation $n(x,k) = n_{\rm LDA}(x,k)$ in the initial state. An estimate of the ground-state propagator in real-space is \cite{Wigner_oscillations_A}
\begin{equation}
	\left< \Psi_{\rm F}^\dagger(x) \Psi_{\rm F}(x') \right> = \frac{\exp \left(-i \int_{x'}^x k_{\rm F}(u) du \right)}{-2\pi i (x-x')} + {\rm c.c.}
\end{equation}
Expanding the argument of the exponential, one gets
\begin{eqnarray*}
	&&\left< \Psi_{\rm F}^\dagger(x+y/2) \Psi_{\rm F}(x-y/2) \right> \\
    && \; = \frac{\exp \left(-i k_{\rm F}(x) y - i\frac{y^3}{24} k_{\rm F}''(x) \right)}{-2\pi i y} + {\rm c.c.}
\end{eqnarray*}
From this expression, one can calculate the Wigner function, and get a better result than the crude approximation $n_{\rm LDA}(x,k)$. For notational simplicity, we focus on the behavior of $n(x,k)$ for $k$ close to $k_{\rm F}(x)$, which is determined only by the first term in the propagator (so we drop the c.c. term),
\begin{eqnarray}
	\label{eq:deviation_LDA}
\nonumber	n(x,k) &=& \int dy e^{i k y} \left< \Psi_{\rm F}^\dagger(x+y/2) \Psi_{\rm F}(x-y/2) \right>  \\
\nonumber    	&\simeq & \int dy \frac{\exp \left(i ( k-k_{\rm F}) y - i\frac{y^3}{24} k_{\rm F}'' \right)}{-2\pi i y} \\
        &=&  C \int dq \, \Theta(k_{\rm F}-k+q) \, {\rm Ai} \left( C q\right)
\end{eqnarray}
where ${\rm Ai}(.)$ is the Airy function, and $C(x) := (k_{\rm F}''(x)/8)^{-1/3}$ is a new characteristic length scale. The convolution with the Airy function induces oscillations of $n(x,k)$ as a function of $k$, close to the Fermi point $k_{\rm F}(x)$, see Fig. \ref{fig:Wigner_osc}.

This discussion shows that replacing the true initial Wigner function $n(x,k)$ by $n_{\rm LDA}(x,k)$ is a valid approximation only if the typical period of the oscillations is small compared to $k_{\rm F}(x)$, such that, when calculating average values of observables, integration over $k$ in the interval $[-k_{\rm F}, k_{\rm F}]$ kills the oscillatory terms. We thus arrive at the criterion $1/C(x) \ll k_{\rm F}(x)$ for the validity of LDA, or in terms of the local density $\rho(x)$,
\begin{equation}
	\rho(x)^{-1} \ll C(x)  \quad \Rightarrow \quad n(x,k) \simeq n_{\rm LDA}(x,k).
\end{equation}
The deviation of the true Wigner function $n(x,k)$ from the LDA result $n_{\rm LDA} (x,k)$ affects the expectation values of charge densities ${\tt q}[h]$. For the charges $q_j$ associated to $h(k) = k^j$, the direct calculation from Eq. (\ref{eq:deviation_LDA}), including the c.c. term, shows that
\begin{eqnarray*}
	{\tt q}_j (x) &=& \int \frac{dk}{2\pi} n(x,k) k^j \\
    &=& \frac{i^j}{2\pi} \left( \frac{\partial^j}{\partial y^j}\frac{e^{-i k_{\rm F} y - i \frac{y^3}{3 C^3}}-{\rm c.c.} }{-i y} \right)_{y=0} .
\end{eqnarray*}
For the first few $j$'s, this gives
\begin{eqnarray*}
	{\tt q}_0(x) &=& \frac{k_{\rm F}(x)}{\pi} \\
	{\tt q}_1(x) &=& 0 \\
	{\tt q}_2(x) &=& \frac{k_{\rm F}(x)^3}{3\pi} - \frac{2}{3 \pi C(x)^3} \\
	{\tt q}_3(x) &=& 0 \\
	{\tt q}_4(x) &=& \frac{k_{\rm F}(x)^5}{5\pi}- \frac{4 k_{\rm F}(x)^2}{\pi C(x)^3} \\
	{\tt q}_5(x) &=& 0 \\
	{\tt q}_6(x) &=& \frac{k_{\rm F}(x)^7}{7\pi}- \frac{10 k_{\rm F}(x)^4}{\pi C(x)^3} + \frac{40 k_{\rm F}(x)}{\pi C(x)^6}
\end{eqnarray*}
Interestingly, since $1/C(x)^3$ is a total derivative, when we integrate over $x$, we see that the expectation values of the total charges $\left< Q_j \right> = \int dx {\tt q}_j(x)$ remain unaffected by the oscillations of the Wigner function for $j \leq 3$. At the level of the expectation values of total charges, the deviation from the LDA result is seen only for $j \geq 4$.

\section{The hydrodynamic regime}
\noindent {\bf\em The hydrodynamic regime.}\quad In classical systems, one assumes that local entropy maximization has occurred because of the fast motion of classical particles and small mean free path (separation of scales). Common wisdom then suggests that hydrodynamics emerges at large times. However, in quantum systems, the eigenstate thermalization hypothesis suggests that near-homogeneous, near-steady states are well approximated, from the viewpoint of local observables, by (generalized) Gibbs ensembles, and thus their dynamics by GHD. In effect, ``molecular chaos" is already present within homogeneous and stationary states. One would therefore expect that time evolution from near-homogeneous, near-steady initial states is instantaneously well represented by GHD. It is possible to suggest, from physical intuition, quantitative criteria. Consider a state with density $\rho(x)$. The total variation, over an interval $d$, of the inter-particle length $l_{\rm int}(x) = 1/\rho(x)$ is $\Delta(x,d) = {\rm max}(|l_{\rm int}(x+y) - l_{\rm int}(x)|:y\in[0,d])$.
On one hand, it is natural to require the relative total variation over $d$ to be small whenever $l_{\rm int}(x)$ is a significant proportion of $d$. That is, $\Delta(x,d)/l_{\rm int}(x)$ times $l_{\rm int}(x)/d$ is $\ll 1$; taking $d\to0$, this is boils down to $|\p_x l_{\rm int}(x)|\ll 1$, or equivalently $l_{\rm int}(x)\ll \rho(x)/|\p_x\rho(x)|$. On the other hand, the scattering length $\varphi(\theta)/p'(\theta)$ should also be small. It is velocity dependent, and an upper bound is  $l_{\rm scat}=\max(\varphi(\theta)/p'(\theta):\theta\in\R)=2/(mc)$. Scattering will occur in a relatively homogeneous region if $\Delta(x,l_{\rm scat})/l_{\rm int}\ll 1$, which boils down to $l_{\rm scat}\ll \rho(x)/|\p_x\rho(x)|$.  Thus, sufficient conditions for the validity of GHD should be
\begin{equation}\label{condihydro}
\max\{l_{\rm int}(x),2/(mc)\}\ll \frac{\rho(x)}{|\p_x\rho(x)|}.
\end{equation}

\section{The NRG-TSA-ABACUS algorithm}
In order to describe the behavior of the Bose gas pre- and post-quench, we must deal with two problems: one, first finding the ground state of the gas in a one-body Gaussian potential, and then two, determining its post-quench evolution in time.  As we will see, it is only the first that is non-trivial.  

To find the pre-quench ground
state, we employ the
truncated spectrum approach, an approach first introduced by V. Yurov and
A. Zamolodchikov to study simple perturbations of the conformal minimal models \cite{TCSA_A}.  This approach in general enables the study of models of
the form,
\begin{equation}
H = H_{integrable}+V_{pert},
\end{equation}
where $H_{integrable}$ will for us be the Lieb-Liniger model and $V_{pert}$ the one-body potential,
\begin{equation}
V_{pert} = V_0 \int^R_0 dx e^{-a x^2}\hat\rho (x),
\end{equation}
where $\hat\rho(x)$ is the density operator.  The method employs the eigenstates of $H_{integrable}$ as a computational basis.  The method
presumes that we have complete control of this basis, knowing both the energies of the eigenstates as well the matrix elements of any relevant operator with respect to this basis.  And in fact we do as the spectrum of the Lieb-Liniger as well as matrix elements of the density
operator are computable.  Because we are using the eigenstates of an interacting theory as a starting point for our numerics,
we, at the very start, incorporate strong correlations into the problem.  This, in principle, alleviates the numerical burden of solving the full
Hamiltonian, at least in comparison of starting with a non-interacting basis.  For a comprehensive review of truncated spectrum methods, see
Ref. \cite{tsa_review_A}.

The essential idea behind the truncated spectrum approach is to order the eigenstates of $H_{integrable}$ by importance
as determined by some metric.  In its simplest form, this metric is their unperturbed energy.  This (infinite) list of states is
then truncated, keeping the most (N, say) important states (as measured by the metric).  This list of states, $\{|E_i\rangle\}^N_{i=1}$,
where state, $|E_i\rangle$, has energy, $E_i$, is then used to form the Hamiltonian matrix,
\begin{widetext}
\begin{equation}
H_N = \begin{bmatrix}
E_1 +\langle E_1|V_{\rm pert}|E_1\rangle  & \langle E_1|V_{\rm pert}|E_2\rangle  & \langle E_1|V_{\rm pert}|E_3\rangle  & \dots & \langle E_1|V_{\rm pert}|E_N\rangle  \\
\langle E_2|V_{\rm pert}|E_1\rangle  &  E_2 +\langle E_2|V_{\rm pert}|E_2\rangle  & \langle E_2|V_{\rm pert}|E_3\rangle & \dots & \langle E_2|V_{\rm pert}|E_N\rangle \\
\langle E_3|V_{\rm pert}|E_1\rangle  &  \langle E_3|V_{\rm pert}|E_2\rangle  & E_3 + \langle E_3|V_{\rm pert}|E_3\rangle & \dots & \langle E_3|V_{\rm pert}|E_N\rangle \\
\vdots & \vdots & \vdots & \ddots \\
\langle E_N|V_{\rm pert}|E_1\rangle  &  \langle E_N|V_{\rm pert}|E_2\rangle  & \langle E_N|V_{\rm pert}|E_3\rangle & \dots & E_N + \langle E_N|V_{\rm pert}|E_N\rangle \\
\end{bmatrix}
\end{equation}
\end{widetext}
This finite dimensional matrix can then be easily diagonalized and both the spectrum and matrix elements of the full theory
determined.

To compute the energies, $E_i$, as well as the matrix elements, $\langle E_i|V_{\rm pert}|E_j\rangle$, for our perturbed Lieb-Liniger model, we employ ABACUS \cite{abacus_A}.  ABACUS is a software
package that enables the efficient computation of these quantities.  These are non-trivial because the determination of energy
of an eigenstate, $|E_i\rangle$, involving $N$ particles necessitates the solution of N-coupled non-linear equations.  These are known as the Bethe ansatz
equations:
\begin{eqnarray}
e^{ip_i L} = \prod^N_{j\neq i} \frac{p_i - p_j + ic/2}{p_i - p_j - ic/2}.
\end{eqnarray}
Here $p_i$ is the momentum associated with the $i-$th particle.  To determine 
the matrix element $\langle E_i,\{p_ik\}^N_{k=1}|V_{\rm pert}|E_j,\{p_jk\}^N_{k=1}\rangle$, ABACUS uses a determinantal expression
for this quantity first developed by N. Slavov \cite{slavnov_A}.

The use of a simple truncation of the spectrum works exceedingly well for perturbations of simple conformal minimal models.  However
for more complicated models, there are strong truncation effects.  To lessen these effects, a variety of strategies are available, both analytical \cite{tcsa_anal_rg_A} and numerical \cite{konik_A}.  The approach we employ here is to adapt the numerical renormalization group \cite{nrg_A} first developed by Kenneth Wilson for the study of quantum impurity problems.  The basic idea is to perform a set of iterative exact diagonalizations where the most important states
are taken into account in the first set of diagonalizations and successively less important states are taken into account in later diagonalizations.  In this way the truncation level of the eigenspace of $H_{Lieb-Liniger}$ can be made much greater.  Rather than employ a basis of $\sim 10^4$
states, one can operate with bases of sizes on the order of $\sim 10^5-10^6$.

The metric that determines which states in the computational basis are important need not be energy.  In fact for the perturbed Lieb-Liniger
model, the energy is not the most robust of metrics.  We have instead developed an adaptive metric that uses the magnitude of matrix
elements of $V_{pert}$ of an unperturbed state with the ground state of the full model (as computed using a small exact diagonalization) \cite{caux_konik_A,scnt_A}.
This metric is much more efficient at determining which unperturbed states will be important for the full problem and enables 
the equivalent study of bases with $\sim 10^6-10^7$ states.

Using this approach we then arrive at the ground state of the gas in the Gaussian potential in the form
\begin{equation}\label{psi_gs}
|\psi_{gs}\rangle = \sum_{i} c_i |E_i\rangle.
\end{equation}
We are confident that this representation is an accurate one for several reasons.  For our purposes here we did not need to consider a very strong Gaussian potential, nominally because we wished to mimic the zero temperature hydrodynamic limit.  In past work, i.e. Refs. \cite{caux_konik_A,kam_A}, we considered stronger one-body potentials with larger number of particles
and were able there to establish the robustness of the technique.   Furthermore we are solely interested in the computing the ground state of the gas in the one-body potential.  If we were in a position to need excited states, we would have to be more cautious.

Having the ground state in hand, Eqn. \ref{psi_gs}, we can then turn to the second problem, determining it's post-quench evolution in time.  This in fact is trivial because the post-quench Hamiltonian is the Lieb-Liniger model itself.  Thus the time evolved state, $|\psi(t)\rangle$, is simply
\begin{equation}
|\psi_{gs}(t)\rangle = \sum_{i} c_ie^{-iE_it} |E_i\rangle,
\end{equation}
that is, because our computational basis is equal to the post-quench
eigenbasis, the coefficients of expansion simply pick up phase factors that depend on the energies, $\{E_i\}$.  It is worthwhile to stress that because we can compute $E_i$ to arbitrary accuracy, we can then evolve the state to arbitrary times without incurring phase errors.

While it is more difficult, we can also consider quenches where we quench from one one-body potential to another non-trivial potential.  In such a case, we must compute using our NRG-TSA-ABACUS algorithm a large number of excited states post-quench so that we can expand the pre-quench ground state into the post-quench eigenbasis.  This is however a tractable problem as demonstrated in Ref. \cite{kam_A} where quenches from parabolic to cosine potentials were studied.  However it does place limits on the strength of the quench as well as the number of particles that can be treated.

\section{Additional graph}
\begin{figure}[ht]
	\begin{center}
    	\includegraphics[width=8.0cm]{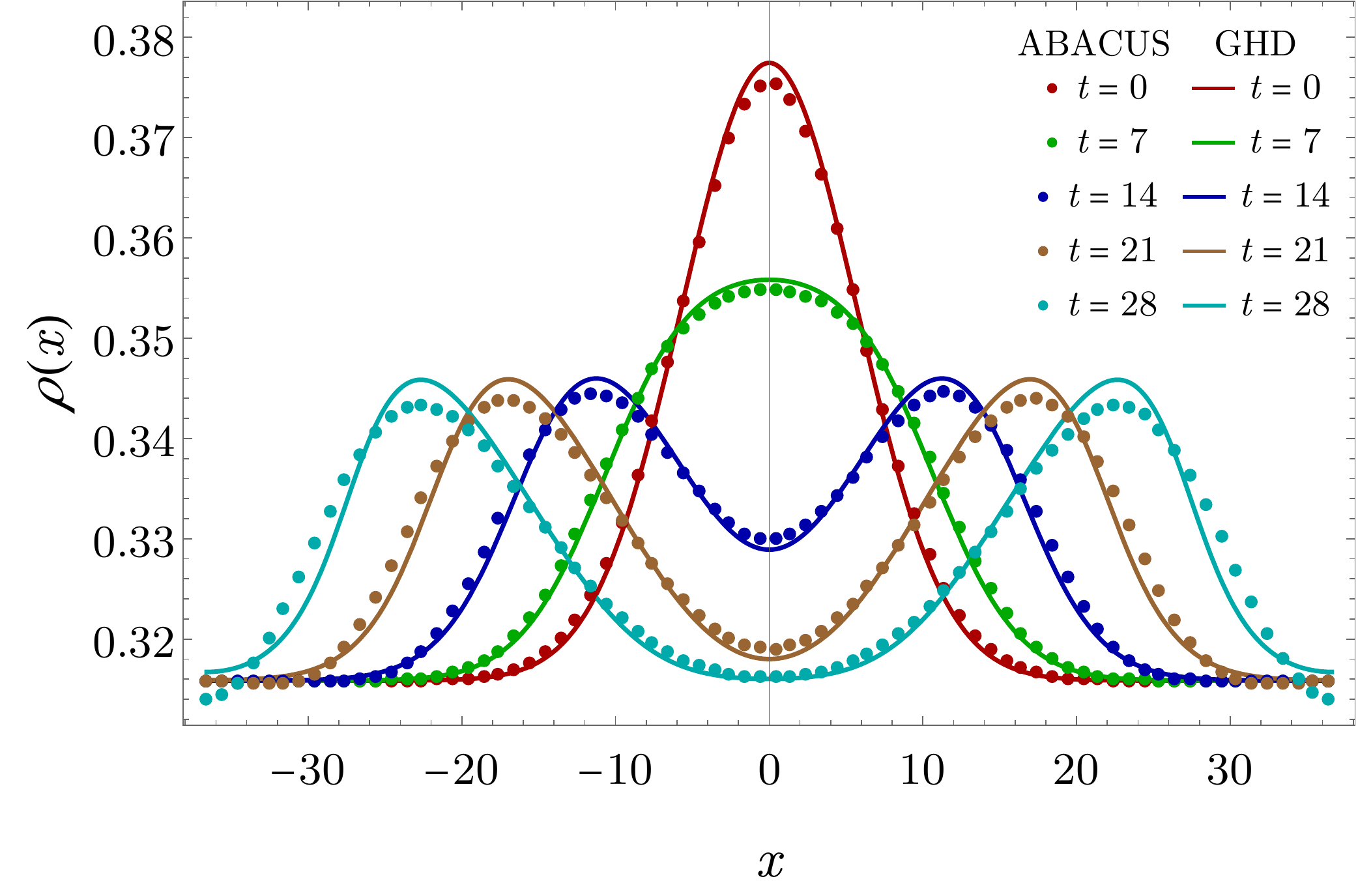}
        \caption{The density profile under the evolution with $2m=1$, $c=1$, $U=0.05$, $a=1/64$, and a choice of $\mu_\infty$ such that
there are $N = 24$ particles in the ground state. The agreement between the initial density obtained by LDA and NRG-TSA-ABACUS numerics is not good in the first place, and the disagreement grows as time is evolved.}
       \label{f3}
    \end{center}
\end{figure}

In the main text, we showed a graph for curves with $\gamma$ being of order 1, which is an intermediate regime. It is then natural to ask how curves would look like in two extreme regimes: the TG regime ($\gamma \gg 1$) and the Gross-Pitaevskii regime ($\gamma \ll 1$). It turns out, however, that the qualitative behavior basically does not differ whatever the coupling constant is. Namely, with any coupling constant, an initial density accumulation splits into two bumps and each bump moves towards right/left with a same speed. Nonetheless, it is instructive to see how a density wave evolves when a trap potential is not smooth enough.

For this purpose, we use a LL gas with $c=1$ as in Fig.\ref{f1}, and a potential $V(x)=-\mu_\infty-Ue^{-ax^2}$ with $U=0.05$, $a=1/64$, and $\mu_\infty$ is determined so that $N=24$. In Fig.\ref{f3} we observe that the initial density prepared by LDA is already not agreeing with NRG-TSA-ABACUS numerics, that is, the initial state is not in the hydrodynamic regime defined by \eqref{condihydro}. This discrepancy is then amplified as time goes, and we also see that the height of the dots by NRG-TSA-ABACUS numerics is lowering. We expect that these effects are generated by the neglected higher conserved charges in preparing the LDA.

\section{Shock dissolution of 2HD into 4HD: a comparison with the classical flea gas}

\begin{figure}[ht]
	\begin{center}
    	\includegraphics[width=8.cm]{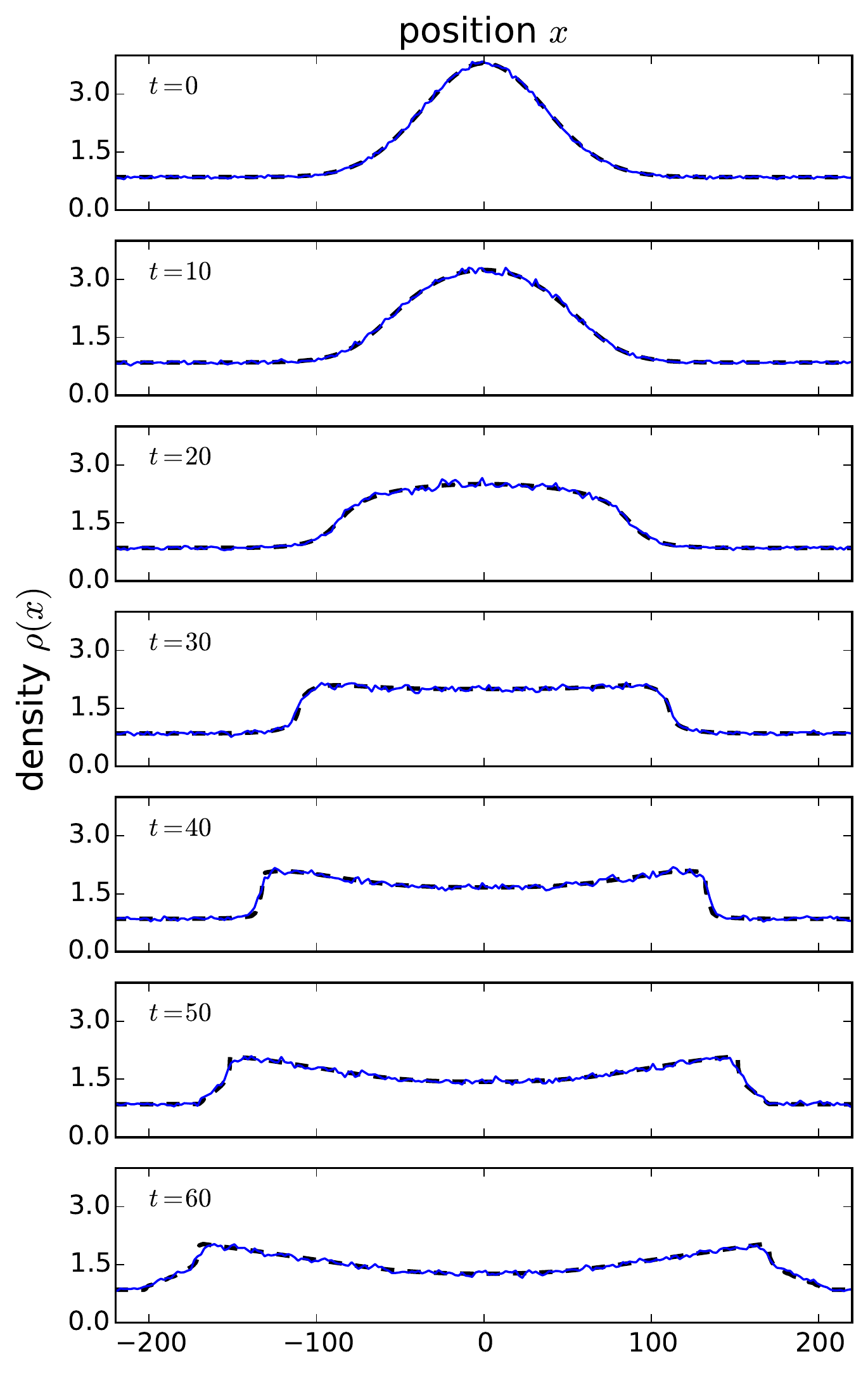}
        \caption{Evolution of the density profile under the evolution with the same parameters as in Fig. \ref{fig:shock_2HD_4HD}: $m=1, c=2$ and an initial potential $V(x) = -5 e^{-(\frac{x}{50})^2}-1$. We compare the data from Fig. \ref{fig:shock_2HD_4HD} in the main text (here in black, dashed) to the classical flea gas (in blue).}
       \label{fig:flea_gas}
    \end{center}
\end{figure}

The mechanism of shock-dissolution elaborated in the main text can be further supported by a comparison with a molecular dynamics of GHD (classical flea gas) \cite{flea_A}. The flea gas algorithm was recently developed by making use of a surprising equivalence between dynamics of the hard rod-like model and GHD. A protocol with the same parameters as in Fig. \ref{fig:shock_2HD_4HD} in the main text is compared against a simulation using the flea gas in Fig. \ref{fig:flea_gas}, showing a perfect agreement. This confirms, in a nontrivial gas (although classical instead of quantum), that the shock dissolution mechanism proposed in the main text is correct.

\section{Sketch of the algorithm for zero-entropy GHD}

In this section, we briefly sketch the algorithm that we use to solve zero-entropy GHD. It is the one we used, for instance, to produce the data plotted in Fig. 1.a in the main text.

The idea is to work with a finite set of curves in the $(x,\theta)$ plane, each of these curves being a set of points where the occupation function $n(x,\theta)$ jumps from $0$ to $1$. We call $\Gamma_{j}$, $j = 1, \dots, p$ these different curves. For example, in Fig. 1.a, there are $p=2$ curves, and they are both plotted in red.

Then we view the zero-GHD entropy equation (Eq. (9) in the main text) as an evolution equation for this set of curves. Namely, if $( x_j , \theta_j)$ is a point on the curve $\Gamma_j$, then after a small time $\delta t$, its new position is $( x_j + \delta x_j, \theta_j)$
where
\begin{equation}
	\label{eq:deltat}
	\delta x_j = v^{\rm eff}_{\{\theta \}} (\theta_j) \delta t .
\end{equation}
The velocity $v^{\rm eff}_{\{\theta \}} (\theta_j)$ is calculated from the set $\{\theta \}$ of all points that lie at the intersection between the vertical line passing through $(x_j, \theta_j)$ and all the curves $\Gamma_{j'}$. It is obtained by solving the integral equation given in the main text, in the paragraph following Eq. (9).

This leads to the following algorithm. Numerically, each curve $\Gamma_j$ is encoded as the interpolation of some discrete set of points $(x_{j,a}, \theta_{j,a})$ for $a=1,2, \dots $. The most basic version of the algorithm uses linear interpolation, but one can also use more refined interpolation schemes such as splines. To go from the configuration at time $t$ to the configuration at time $t+ \delta t$, we do a loop over all the points labelled by $(j,a)$. For each point:
\begin{itemize}
\item we start by finding all the intersections between the vertical line at $x = x_{j,a}$ and all the curves $\Gamma_{j'}$ for $j' = 1,\dots, p$. This gives us a set $\{\theta \}$.
\item we order the elements of the set $\{\theta \}$ and label them as $\theta_1^- < \theta_1^+ < \theta_2^- < \dots < \theta_k^- < \theta_k^+$.
\item we calculate ${\rm id}^{\rm dr}_{\{ \theta \}} (\theta_{j,a})$ and $1^{\rm dr}_{\{ \theta \}} (\theta_{j,a})$. To do this, one needs to solve the linear integral problem $f^{\rm dr}_{\{ \theta \}}  (\alpha) = f(\alpha) + \sum_{q=1}^k  \int_{\theta_q^-}^{\theta_q^+} d \gamma \varphi(\alpha-\gamma) f^{\rm dr}_{\{ \theta \}}  (\gamma)$. This is done by discretizing the integral, which leads to a matrix formulation of the problem, of the form $f^{\rm dr} = f + M \cdot f^{\rm dr}$ where $f$ and $f^{\rm dr}$ are finite-dimensional vectors.
\item we evaluate the ratio $v^{\rm eff}_{\{ \theta \}} (\theta_{j,a}) = {\rm id}^{\rm dr}_{\{ \theta \}} (\theta_{j,a})/ 1^{\rm dr}_{\{ \theta \}} (\theta_{j,a})$, and use it to calculate the new position of the point $(x_{j,a} + \delta x_{j,a},\theta_{j,a})$ after $\delta t$, according to Eq. (\ref{eq:deltat}).
\end{itemize}
We find that the algorithm performs better if one  reparametrizes the curves $\Gamma_j$ from time to time.  Namely, it is possible that the distance between successive points $(x_{j,a},\theta_{j,a})$ and $(x_{j,a+1},\theta_{j,a+1})$ increases during the evolution, so that the initial discrete set of points does not provide a good description of the curve $\Gamma_j$ any longer. To avoid that, it is useful to chose a new discrete set points on the curve $\Gamma_j$ every once in a while, and then carry on with the evolution.

\bibliographystyle{alpha}

\end{document}